\documentclass[
]{ceurart}

\usepackage{numprint}
\usepackage{enumitem}
\usepackage{tikz}
\usepackage{pgfplots}
\usepackage{pgfplotstable}
\usepackage{csvsimple}
\pgfplotsset{compat=newest}

\begin{document}

\copyrightyear{2022}
\copyrightclause{Copyright for this paper by its authors.
  Use permitted under Creative Commons License Attribution 4.0
  International (CC BY 4.0).}

\conference{CICM 2021: 14th Conference on Intelligent Computer Mathematics,
  July 26--31, 2021, Timisoara, Romania}

\title{Structure in Theorem Proving: Analyzing and
Improving the Isabelle Archive of Formal Proofs}

\author{Fabian Huch}[
orcid=0000-0002-9418-1580,
email = huch@in.tum.de,
url=https://www21.in.tum.de/home/~huch
]
\address{Technische Universität München, Boltzmannstraße 3, 85748 Garching, Germany}


\begin{abstract}
The Isabelle Archive of Formal Proofs has grown to a significant size in the past years.
It makes up for an impressive body of research,
which enables a number of statistical approaches to various aspects in theorem proving,
and has not yet been utilized exhaustively.
However, the growing size also poses some challenges to address:
Material becomes increasingly harder to find,
reusability and ease of understanding become more important.
This thesis abstract summarizes my research plans on those topics and briefly touches on preliminary results,
which indicate that the node in-degree of the dependency graph of the archive follows a scale-free distribution.
\end{abstract}


\maketitle
\section{Introduction}
\emph{Isabelle} is an interactive theorem prover \cite{Isabelle1998Paulson} with
a large collection of formalized material, the \emph{Archive of Formal Proofs} (AFP).
At the time of writing, the AFP consists of nearly three million lines of code,
and more than \numprint{167000} lemmas have been proven
in its close to \numprint{600} different entries \cite{Statistics2021Afp}.

The entities defined by those theories and their relationships can be described as a dependency graph 
(sometimes also referred to as \emph{General Dependency Network} \cite{ClusteringMetrics2014Milos}).
For software systems, dependency graphs
have been found to exhibit certain structural properties. 
For instance, their in-degree ($k_{in})$ distribution is often \emph{scale-free},
i.e. it follows some power law $\Pr(k_{in})\propto k_{in}^{-\gamma}$ \cite{Networks2003Myers},
in contrast to random graphs where each possible edge is present with a fixed probability (Erdős–Rény model).
Graphs with such topological properties are called \emph{complex networks},
which have been studied in the context of real-world graphs from many different areas \cite{ComplexNetworks2002Sole}.
In contrast to research focused on object-oriented characteristics (e.g., classes and inheritance),
structural analysis on dependency graphs of software systems can be transferred to the field of formal theories more directly.

Another important application of this dependency network is proof automation.
Data mining methods to extract patterns from graphs have been extensively studied,
for instance \emph{frequent itemset mining} to find subsets of nodes that frequently occur together \cite{MiningGraph2010Aggarwal,MiningGraph2006Cook}.
The AFP provides enough data to enable pattern mining for proofs as well as evaluate the effectiveness of derived automation.

\section{Related Work}\label{sec:relwork}

Regarding the AFP, empirical analysis has already been done by \citeauthor{MiningAFP2015Blanchette} in \cite{MiningAFP2015Blanchette}.
They measured the AFP size and authorship distribution and how those evolved over time,
computed the session graph, proof depths, and compared proof size to definition size and statement complexity (by number of clauses).
Moreover, they benchmarked how many theorems \emph{sledgehammer} could solve on a representative test suite.
This research provides a good baseline of empirical data;
I plan to follow up on it by analyzing the underlying GDN of AFP theories, which can possibly yield some deeper results.

Very recently, usability of the AFP has also gotten some attention.
Two different search engines for theory contents were introduced:
The \emph{FindFacts} search by \citeauthor{FindFacts2020Huch} in \cite{FindFacts2020Huch},
and the concept-oriented \emph{SErAPIS} engine by \citeauthor{Serapis2020Stathopoulos} in \cite{Serapis2020Stathopoulos}.
Additionally, \citeauthor{AfpEval2021Mackenzie} evaluated usability of the AFP webpage in \cite{AfpEval2021Mackenzie} and built the prototype for a re-design.

Complex networks for software systems were first introduced by \citeauthor{Networks2003Myers} in \cite{Networks2003Myers}.
In \cite{DefectsMetrics2008Zimmermann}, \citeauthor{DefectsMetrics2008Zimmermann} measured the correlation between software defects and a comprehensive list of network metrics;
\citeauthor{Networks2012Subelj} later surveyed quality indicators,
and performed package prediction by clustering on the network \cite{Networks2012Subelj}.
While there is no concern about defects in theorem proving due to the nature of the field,
my aim is to transfer some of those findings to formal theories
in order to detect structural problems that impair reusability and readability.

For proof automation, the idea of learning from existing material has been around for some time.
In Isabelle, \citeauthor{MiningTactics2007Duncan} derived tactics from proofs script sequences using Markov models and genetic programming with the goal of full automatization in \cite{MiningTactics2007Duncan}, which was successful for less complex lemmas.
More recently, \citeauthor{ProofMining2020Nawaz} used high utility itemset mining on a syntactic level to discover patterns in proofs scripts of \emph{PVS} \cite{ProofMining2020Nawaz}.
In contrast, I aim to extract patterns from the more fine-grained theorem dependency graph.
This also makes it possible to utilize structured Isar proofs in the pattern extraction.

\section{Research Topics}
In the following, my research plans are separated into concrete questions and topics that involve engineering tasks.

\subsection{Research Questions}

\begin{enumerate}[label=\bfseries{RQ\arabic*}:, leftmargin=*]
    \item Do theory dependency networks follow the topological patterns typically found in complex networks?
    \item Which metrics are relevant quality indicators for formal theories? Can they expose structural problems?
    \item Is it possible to detect elements that need to be refactored? Can theory- and session-structure be predicted to generate refactoring recommendations?
    \item How can visualization be helpful to better understand the structure of formalizations?
    \item Which patterns can we learn from theorem networks? Can they be used to improve automation?
\end{enumerate}

\subsection{Further Topics}
There are several aspects I want to improve on in the AFP.
By introducing digital object identifiers,
entries can be much better referenced.
Identifiers could potentially even be assigned to individual concepts of an entry,
though creating such a mapping poses lots of challenges.
Moreover, introducing more fields for entries,
such as subject classification and links to (print-)publications,
can help organizing the AFP in the future.
Finally, the AFP website needs to be improved for better usability -- there is already a prototype (as discussed in \autoref{sec:relwork}), which still needs to be integrated.

Currently, the AFP does not allow proofs by \texttt{sorry} (which are admitted with the help of an oracle \cite{IsarRef2021Wenzel}).
However, it is often desirable to have what is sometimes referred to as \emph{Formal Abstract} \cite{FormalAbstract2017Hales}:
a formalization of results from literature without rigorous proofs.
To that end, I want do add a \emph{proof by reference} mechanism,
which weakly checks references and allows such proofs to be added to the AFP in a separate, less trusted, session group.

In Isabelle, I plan to integrate the FindFacts search into the prover IDE (and add support for type-classes and locales),
as well as tooling for clone detection and advanced IDE functionality such as structural search and replace.

\section{Preliminary Results}
\begin{figure}[!htbp]
    \centering
    \begin{tikzpicture}
    \begin{axis}[xlabel=$k_{in}$, ylabel=$\Pr(k_{in})$, xmode=log,ymode=log,grid=major]
    \addplot[only marks,black,opacity=0.6,mark=o] table[x=deg,y=count,col sep=comma] {data.csv};
    \end{axis}
    \end{tikzpicture}
    \caption{In-Degree distribution the dependency graph from Isabelle and AFP theories (release $2021$), on a log-log scale.}
    \label{fig:indegdist}
\end{figure}
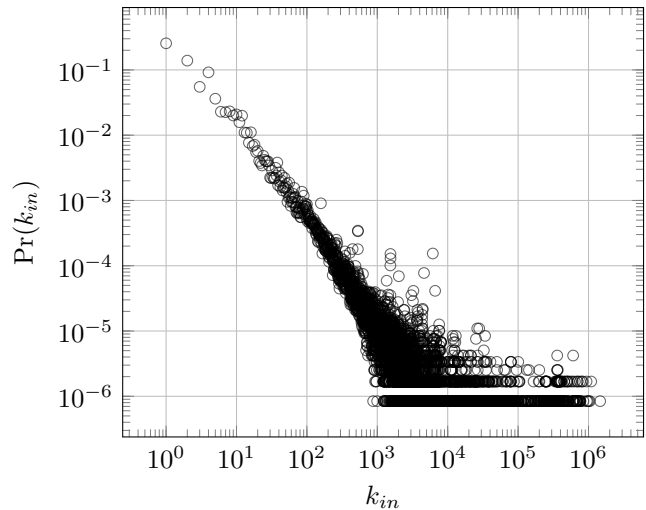

The dependency graph of Isabelle and the AFP (release 2021) consists of approximately $2.1\cdot 10^6$  nodes and $2.5\cdot 10^8$ directed edges (when considering all types of entities and relationships).
\autoref{fig:indegdist} shows the in-degree distribution of that graph.
For $k_{in}>2$, it is clearly linear on the $\log$-$\log$ scale;
this indicates a power-law distribution, which is typical for scale-free networks \cite{ComplexNetworks2002Sole}.
The individual AFP components show similar characteristics;
this illustrates why complex network science is relevant to the topic.

%%
%% Define the bibliography file to be used
\bibliography{library}


\end{document}